\documentclass[reprint,aps,prl]{revtex4-2}
\usepackage{amsmath,amssymb,amsfonts,dsfont}
\usepackage{graphicx}
\usepackage{color}
\usepackage{hyperref}
\usepackage{subfigure}
\usepackage{dcolumn}% Align table columns on decimal point
\usepackage{bm}% bold math
\usepackage{float}
\usepackage{mathtools}
\usepackage{amsthm}
\usepackage{bbm}
\usepackage{pxfonts}
\usepackage{pstricks}
\usepackage{verbatim}
\usepackage[normalem]{ulem}

\definecolor{orange}{cmyk}{0,0.5,1,0}
\definecolor{rossoCP3}{cmyk}{0,.88,.77,.40}
\definecolor{graa}{rgb}{0.8,0.8,0.8}
\definecolor{blaa}{rgb}{0.2,0.2,0.6}
\hypersetup{
	colorlinks, 
	bookmarksopen, 
	bookmarksnumbered,
	citecolor=blaa, 		%color of links to bibliography
	linkcolor=rossoCP3,	%color of internal links
	urlcolor=rossoCP3,			%color of external links 
	    }
\newcommand{\beq}{\begin{eqnarray}}
\newcommand{\eeq}{\end{eqnarray}}

\newcommand{\SU}{\mathrm{SU}}

\newcommand{\U}{\mathrm{U}}

\newcommand{\Tr}{\text{Tr}}

\begin{document}
\noindent\mbox{}\hfill{\small
 IPPP/25/65}\par\vspace{2ex}
\title{The Good Qualities of the Weak Axion
%\\ ~ \\ \& \\ ~\\ The $B+L$ ALP Generalization \\ ~ \\ \& \\ ~\\
%Weak Axion, Strong Limits
}

\author{Giacomo Cacciapaglia}
\email{cacciapa@lpthe.jussieu.fr}
\affiliation{Laboratoire de Physique Theorique et Hautes Energies {\color{rossoCP3}LPTHE}, UMR 7589, Sorbonne Universit\'e \& CNRS, 4 place Jussieu, 75252 Paris Cedex 05, France.}
\author{Francesco Sannino}
\email{sannino@qtc.sdu.dk}
\affiliation{{\color{rossoCP3}{$\hbar$}QTC} \& the Danish Institute for Advanced Study {\color{rossoCP3}\rm{Danish IAS}},  University of Southern Denmark, Campusvej 55, DK-5230 Odense M, Denmark;}
\affiliation{Dept. of Physics E. Pancini, Universit\`a di Napoli Federico II, via Cintia, 80126 Napoli, Italy;}
\affiliation{INFN sezione di Napoli, via Cintia, 80126 Napoli, Italy}
%\affiliation{Scuola Superiore Meridionale, Largo S. Marcellino, 10, 80138 Napoli, Italy,}
\author{Jessica Turner}
\email{jessica.turner@durham.ac.uk}
\affiliation{Institute for Particle Physics Phenomenology, Durham University, South Road, DH1 3LE,
Durham, United Kingdom}
%%%%%%%%%%%%%%%%%%%%%%%%%%%%%%%%%%%%%%%%%%%%%%%%%%%%%%%%%%%%%%%%%%%%%%%%%%

\begin{abstract}
The presence of a topological susceptibility in the electroweak sector of the Standard Model motivates the existence of a {\it good quality} weak axion $a_W$, associated with the spontaneous breaking of $B\!+\!L$. Its anomalous couplings and tiny mass, generated from electroweak instantons, render $a_W$ photophobic. We find that the strongest bound on the associated decay constant, $f_W$, stems from a loop-induced coupling to electrons, leading to $f_W \gtrsim 1000$~TeV from stellar cooling. Spontaneous breaking of the abelian ${B\!+\!L}$ symmetry induces proton decay via higher dimensional operators controlled by a new physics scale, $\Lambda$.  Existing Super-Kamiokande limits on these decay channels constrain the new physics scale to be $\Lambda \gtrsim 10^{12}$~GeV. The  characteristic channel $p\to e^+ a_W$ and other possible operators mediating interactions with the Standard Model fields yield signals which are not detectable within the allowed parameter space. 
Future proton decay searches at the next-generation of neutrino experiments offer the most promising avenues to test the {\it good qualities} of the weak axion paradigm.

\end{abstract}

\maketitle
\section{Introduction}
The vacuum structure of non-Abelian Yang--Mills gauge theories is rich and features a non-vanishing topological susceptibility. For this reason, an $\SU(N)$ gauge theory admits a topological $\theta$--term:
\begin{equation}
\label{eq:topoLYM}
\mathcal{L}_\text{YM}\supset \theta\, \epsilon^{\mu\nu\rho\sigma}\,\Tr {\cal F}_{\mu\nu}{\cal F}_{\rho\sigma}\,,
\end{equation}
where ${\cal F}_{\mu\nu}$ is the field strength tensor and the trace acts over the $N^2\!-\!1$ generators. In the presence of fermions, the $\theta$--angle is tied to an anomalous global $\U(1)$ symmetry~\cite{tHooft:1976rip}, and its physical effects manifest through CP violation~\cite{Baluni:1978rf}.

In QCD, this structure underlies the strong CP problem~\cite{Peccei:2006as} and bounds on hadronic electric dipole moments~\cite{Crewther:1979pi} require an unnaturally small effective $\theta$. The Peccei--Quinn (PQ) mechanism~\cite{Peccei:1977hh,Peccei:1977ur} dynamically relaxes $\theta$ via a spontaneously broken anomalous $\U(1)_{PQ}$, predicting a light pseudoscalar, the axion. In this sense, $\U(1)_{PQ}$ is a \emph{good quality} symmetry: explicit breaking by renormalizable couplings is absent, and residual violations (e.g. from gravity \cite{Georgi:1981pu}) must be sufficiently suppressed \cite{Kamionkowski:1992mf,Barr:1984qx,Holman:1992us}. In contrast, the axial $\U(1)_A$ is a \emph{bad quality} symmetry as it is explicitly broken by quark masses; it is further spontaneously broken by the quark condensate leading to a would-be  Nambu-Goldstone $\eta'$ that, even with massless quarks, is heavy due to the axial anomaly. For massive quarks, $\theta$ becomes physical and its smallness requires tuning between the QCD vacuum angle and phases associated to the quark mass matrices. Generalizations to different matter representations and multi-sector extensions are reviewed in~\cite{DiVecchia:2013swa}, where also the effective Lagrangian of \cite{Crewther:1979pi}, describing the interactions among baryons and pseudo-Nambu-Goldstone mesons, was corrected.  

An analogous topological structure exists in the electroweak (EW) sector. The weak $\SU(2)_L$ gauge group also admits a $\theta$--term, but within the Standard Model (SM) it is unobservable \cite{Anselm:1992yz,Anselm:1993uj,FileviezPerez:2014xju} as it can be rotated away by linear combinations of baryon number ($B$) and lepton number ($L$) global transformations. More precisely, it is the $B\!+\!L$ combination that has an anomaly with $\SU(2)_L$ affecting the weak vacuum of the SM. This symmetry remains a \emph{good quality} one, as long as $B$ and $L$ violations remain small (typically arising from grand unification \cite{Georgi:1974sy} and Majorana neutrino masses \cite{Weinberg:1979sa}).
Still, the presence of a nontrivial topological susceptibility suggests that a pseudoscalar degree of freedom can saturate the corresponding three-form dynamics~\cite{Dvali:2005an}. In the language of Higgsed three-forms, an anomalous Abelian symmetry gauges a composite three-form that acquires a mass by “eating” a pseudoscalar, ensuring consistency of the infrared description~\cite{DiVecchia:1980yfw,Hebecker:2019vyf}. This logic has motivated the identification of an EW-associated pseudoscalar, sometimes dubbed $\eta_W$~\cite{Dvali:2024zpc}.

In our recent work~\cite{Cacciapaglia:2025xmr} we showed that, in the SM with a \emph{good quality} $B\!+\!L$, the $\eta_W$, associated with the $\SU(2)_L$ three-form, is not an independent propagating singlet: once color confinement and SM gauge invariance are properly enforced, its low-energy manifestation maps onto the  pseudoscalar hydrogen  ground state rather than a new particle. Henceforth, there is no need for a light new degree of freedom that receives a mass only through the anomaly, as argued in \cite{Dvali:2024zpc,Dvali:2025pcx}.

In this work we take a different route and investigate the \emph{good quality weak axion} which we denote by $a_W$. We introduce it as a pseudo-Nambu–Goldstone boson of a spontaneously broken  $\U(1)_{B+L}$, as this is the only good quality global charge in the SM with an irreducible $\SU(2)_L$ gauge anomaly. The weak axion ($a_W$), therefore, couples to EW gauge bosons and fermions, $f$, as follows:
\begin{equation}
\label{eq:EWcouplings}
\begin{aligned}
\mathcal{L}\supset \frac{a_W}{f_W}\!\left[c_W\frac{\alpha_2}{8\pi}\,W^a_{\mu\nu}\tilde W^{a\,\mu\nu}
+ c_B \frac{\alpha_Y}{8\pi}\,B_{\mu\nu}\tilde B^{\mu\nu}\right]
\\
+\; \sum_f c_f\,\frac{\partial_\mu a_W}{f_W}\,\bar f\gamma^\mu\gamma_5 f \, ,
\end{aligned}
\end{equation}
where $\tilde{W} = \frac{1}{2} \epsilon W$ and $\tilde{B} = \frac{1}{2} \epsilon B$, while $f_W$ is the $a_W$ decay constant which we assume is common to the gauge and fermion sectors. 
The anomaly coefficients $c_W$ and $c_B$ can be computed, leading to $c_W = - c_B = 1$ in our conventions. We note that this $a_W$ has tree-level couplings to $WW$, $ZZ$ and $Z\gamma$. However, as
$B$ and $L$ do not have a gauge anomaly with QED, the $a_W$ coupling to photons vanishes as $c_\gamma= c_W + c_B=0$. Nevertheless, a coupling is generated at loop level \cite{Bauer:2017ris}, and it is proportional to the square of the $a_W$ mass, $m_a^2$. If the $a_W$ mass is smaller than all the masses of the SM, as expected in a good quality case where the $a_W$ mass only stems from suppressed weak instantons, the coupling to photons is highly suppressed.
Hence, the $a_W$ is essentially \emph{photophobic}. Couplings to fermions are more model-dependent, and they may be absent at leading order but generated at loop level with a logarithmically enhanced coupling \cite{Bauer:2017ris}. A photophobic axion-like particle (ALP) with such couplings has been studied in \cite{Craig:2018kne}.
In the case of the $a_W$, there might also exist a coupling to SM fermions which stems from the $B\!+\!L$ charge of the source field: the minimal choice would be to couple it to an operator containing three quarks and one lepton, $qqql$ \cite{Dvali:2024zpc}. This coupling generates proton decay processes which we will study in the following section. 

More specifically, we construct an effective model for the $a_W$, and study the related phenomenological constraints. Since we do not commit to a specific $a_W$ the present analysis is general  encompassing also the light $\eta_W$ case investigated in \cite{Dvali:2024zpc,Dvali:2025pcx,Davoudiasl:2025qqv}.

\section{The Weak Axion}
The $a_W$ can be introduced beyond the SM, as mentioned in \cite{Dvali:2024zpc}, as a pseudo-Nambu-Goldstone boson arising from a new scalar field, $\Phi_W$, that carries a global $B\!+\!L$ charge and acquires a vacuum expectation value (VEV):
\begin{equation}
    \Phi_W = \langle \Phi_W \rangle \, e^{\,i\, a_W/(\sqrt{2} f_W)}\,, 
\end{equation}
where $f_W$ is the axion decay constant. The two scales, the VEV and the decay constant, can be related. However, their precise relation depends on the nature of the $a_W$: if $\Phi_W$ is an elementary weakly-coupled scalar, then $\langle \Phi_W \rangle = f_W$. If $a_W$ is generated by a composite sector, then $\langle \Phi_W \rangle > f_W$. We estimate that in such a scenario $\langle \Phi_W \rangle \sim g^\ast f_W$ with $g^\ast \,\sim 4\pi/\sqrt{N}$, $N$ being the number of confining colors.  The n\"aive dimensional analysis and large-$N$ counting stems from having the new fermions transforming according to the fundamental representation of an $SU(N)$ gauge group \cite{tHooft:1973alw,Witten:1979kh}.  If the new fermions  transform according to a generic representation $R$ of the gauge group the scaling becomes $g^\ast \sim 4\pi/\sqrt{d(R)}$ \cite{Sannino:2024xwj}. 

The global $U(1)_{B+L}$ charge of $\Phi_W$ is communicated to the SM via couplings to the quarks and leptons. A minimal gauge and Lorentz invariant portal-like operator consistent with $U(1)_{B+L}$ is
\begin{equation} \label{eq:phiWop}
    \frac{\Phi_W}{\Lambda^3}\,qqql\,,
\end{equation}
which fixes $Q_{B+L}(\Phi_W)=-2$. Here $\Lambda$ denotes a UV scale independent of $f_W$, where the effective field theory validity requires that $\Lambda \gg \langle \Phi_W \rangle$. After spontaneous breaking of $B\!+\!L$,  Eq.~\eqref{eq:phiWop} yields
\begin{equation}
    \frac{\langle \Phi_W\rangle}{\Lambda^3}\,qqql
\;+\;
i\,\frac{a_W}{\sqrt{2}f_W}\,\frac{\langle \Phi_W\rangle}{\Lambda^3}\,qqql
\;+\;\cdots\,,
\end{equation}
where the first term matches the usual dimension-six $qqql$ operator responsible for proton decay. The second term sources a contact vertex for $p\to e^+ a_W$. In the following, we will investigate the phenomenology of these channels.
We note that a lepton portal term, 
$(\Phi_W/\Lambda_l^2)\,(l H)(l H)$,
is also allowed by the $B\!+\!L$ charges and enters one mass dimension lower than Eq.~(\ref{eq:phiWop}).
It would contribute to the Weinberg neutrino mass operator \cite{Weinberg:1979sa} at the price of breaking $B\!-\!L$. Furthermore, as for any new scalar field, a renormalizable portal coupling to the SM Higgs can be added, $\Phi_W^\dagger \Phi_W H^\dagger H$, which is constrained by the Higgs coupling measurements. 

Independently of its origins, the $a_W$ mass is generated by non-perturbative EW instanton effects, leading to a highly suppressed value:
\begin{equation}
    m_{a_W} \sim \kappa\ \frac{v_\text{EW}^2}{f_W} e^{-\pi/\alpha_2} \sim \kappa\frac{ (1~\mbox{TeV})}{f_W} \times 6\cdot 10^{-29}~\mbox{eV}\,,
\end{equation}
where $\kappa$ is an $\mathcal{O}(1)$ coefficient. 
This non-perturbative suppression ensures the mass is far below the EW scale and much smaller than any SM mass (including neutrinos), hence validating $a_W$'s photophobic nature.

The weak axion couples to SM particles similarly to the QCD axion, but without a direct coupling to gluons and photons. The bounds on $f_W$, therefore, follow those studied in \cite{Craig:2018kne} and are dominated by red giant cooling. We will first review such bounds, and then discuss the complementary limits on the portal operators, such as that in Eq.~\eqref{eq:phiWop}. 

\subsection{Weak Axion Photophobia }
For the case of the photophobic axion-like particle \cite{Craig:2018kne}, the $a_W$  has anomaly couplings $c_W=-c_B=1$ and vanishing couplings to fermions at scales $\sim\langle\Phi_W\rangle$, where $U(1)_{B+L}$ is broken. 
Below this scale, radiative corrections generate couplings to all SM fermions \cite{Bauer:2017ris}. 
For the light masses relevant for the $a_W$, the most important such coupling is to electrons: 
\begin{equation} \label{eq:cl}
    \frac{c_{e}}{f_W} = - \frac{3 \alpha^2}{16 \pi^2 f_W} \left[ \frac{3}{4\sin^4 \theta_W} - \frac{Y_{l}^2 + Y_e^2}{\cos^4 \theta_W}\right] \ln \left(\frac{\langle \Phi_W \rangle^2}{m_W^2}\right)\,,
\end{equation}
where $\alpha \sim 1/137$ is the fine-structure constant at the EW scale, $\theta_W$ is the Weinberg angle, while $Y_l=-1/2$ and $Y_e=-1$ are the hypercharges of the SM lepton fields. This coupling is most strongly constrained by red giant cooling \cite{Raffelt:1994ry}, which bounds the effective coupling of electrons to axion-like particles to be $y_e < 2.5\cdot 10^{-13}$. In our model, it is defined as $y_e = (2 m_e c_e)/f_W$. By using the expression in Eq.~\eqref{eq:cl}, we can extract a bound on $f_W$. In the elementary case ($\langle \Phi_W \rangle = f_W$), we find
\begin{equation}
    f_W \gtrsim 1000~\mbox{TeV}\quad \mbox{(red giants)}\,. 
\end{equation}
This bound is larger than the one found in \cite{Davoudiasl:2025qqv}, which did not include the log-enhancement. Due to the log dependence on the VEV, the bound is similar in the composite $a_W$ case. As  stressed above, the coupling to photons is extremely suppressed due to the tiny $a_W$ mass, hence no further bounds arise. We further agree with \cite{Davoudiasl:2025qqv} that by equating $a_W$ with the light $\eta_W$ \cite{Dvali:2024zpc,Dvali:2025pcx} is contrived because current constraints already place $f_W$ well above the electroweak scale. Consistently with \cite{Cacciapaglia:2025xmr}, once confinement is included, $\eta_W$ is not exotic but the CP-odd hydrogen eigenstate.

The $a_W$ could also be directly produced at future colliders via the $Z\gamma$ coupling, leading to the decays of the $Z$ boson into a single photon plus missing energy. At the Tera-Z run of the FCC-ee \cite{FCC:2025lpp} or CEPC \cite{CEPCStudyGroup:2023quu}, some sensitivity could be achieved, as studied in \cite{Polesello:2025gwj,Wang:2025ncc}. However, taking the result from \cite{Wang:2025ncc}, we infer that the collider sensitivity can reach up to $f_W \sim 10$~TeV, hence remaining within the region excluded by red giants.

\subsection{Weak Axion Bounds from  Proton Decay}
In this section, we provide estimates of the new–physics scales  associated with the portals
$(\Phi_W/\Lambda^3)\,qqql$ and $(\Phi_W/\Lambda_l^2)\,(lH)(lH)$. We assume that a single operator dominates and neglect renormalization group (RG) running between the UV-scale and the hadronic scale and effects from nuclear matrix elements. We estimate that RG and nuclear matrix–element effects can shift the inferred scales by factors of a few. 

A distinctive feature of the weak axion, absent for the QCD axion and generic ALPs, is the portal in Eq.~\eqref{eq:phiWop}. Once $\Phi_W$ acquires a VEV, it induces the GUT-like four-fermion operator
\begin{equation}
\label{eq:opGUT}
\frac{\langle \Phi_W\rangle}{\Lambda^3}\, qqql\,,
\end{equation}
which mediates proton decay. The most stringent current bound comes from Super-Kamiokande’s $p\to e^+\pi^0$ search~\cite{Super-Kamiokande:2020wjk}, $\tau_{p\to e^+ \pi^0} \ge 1.4\cdot 10^{34}\,\text{yr}$, implying~\cite{Ohlsson:2023ddi}
\begin{equation}
\sqrt{\frac{\Lambda^3}{\langle \Phi_W \rangle}} \;\ge\; 2.4\times 10^{16}\ \text{GeV}\,.
\end{equation}
Consistency of the EFT requires $f_W \simeq \langle \Phi_W\rangle \le \Lambda$. 
Assuming $\langle \Phi_W\rangle\simeq f_W=10^{6}\,\text{GeV}$ implies $\Lambda \gtrsim 8.3\times 10^{12}\,\text{GeV}$. We note that if $\langle \Phi_W \rangle=4\pi f_W$, then $\Lambda$ increases by a factor around two at most.
Looking ahead, adopting the Hyper-Kamiokande 10-year sensitivity $\tau_{p\to e^+\pi^0}=5\times 10^{34}\,\text{yr}$~\cite{Hyper-Kamiokande:2018ofw}, we obtain the projection
\begin{equation}
\sqrt{\frac{\Lambda^3}{\langle \Phi_W \rangle}} \;\ge\; 3\times 10^{16}\,\text{GeV}\,,
\end{equation}
which for the scenario $\langle \Phi_W \rangle =f_W = 10^6$~GeV, implies $\Lambda \gtrsim 9.7\times 10^{12}$ GeV. 

A more direct signature arises from the first term in the expansion of Eq.~\eqref{eq:phiWop} in the $a_W$ field, which produces a direct coupling of $a_W$ to proton and electron:
\begin{equation}
\frac{\langle \Phi_W \rangle}{\sqrt{2} f_W}\,\frac{a_W}{\Lambda^3}\,qqql\;\sim\;
\frac{\langle \Phi_W \rangle}{\sqrt{2} f_W}\,\frac{\Lambda_{\rm QCD}^3}{\Lambda^3}\,
a_W\,\bar{\psi}_e \psi_p\,,
\end{equation}
\begin{figure}[t!]
    \centering
\includegraphics[width=0.9\linewidth]{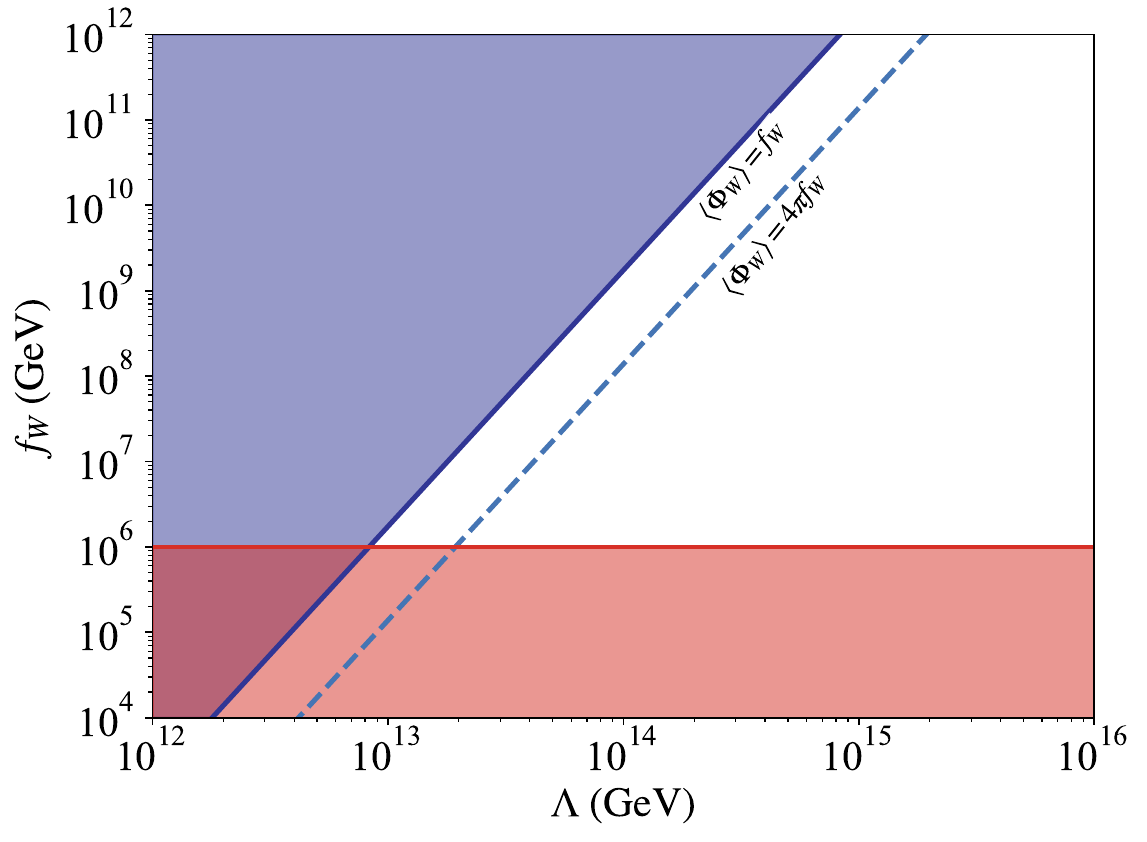}
    \caption{Exclusion limits on the $f_W$--$\Lambda$ plane from red giant cooling (red) and proton decay (blue) from Super-Kamiokande. The solid lines correspond to the elementary case ($\langle\Phi_W\rangle=f_W$) and the dashed one to the composite one ($\langle\Phi_W\rangle=4\pi f_W$).}
    \label{fig:aabound}
\end{figure} which induces the proton decay $p\to e^+ a_W$. The resulting Yukawa-type coupling gives
\begin{equation}
\Gamma_{p\to e^+ a_W}
= \frac{m_p}{8\pi}\,
\frac{\Lambda_{\rm QCD}^6}{\Lambda^6}\,
\frac{\langle \Phi_W \rangle^2}{f_W^2}\,.
\end{equation}
Super-Kamiokande’s search for a (quasi-)massless axion yields
$\tau_{p\to e^+ a_W} \ge 7.9\times 10^{32}\,\text{yr}$~\cite{Super-Kamiokande:2015pys}. 
This implies:
\begin{equation}
\Lambda \left( \frac{f_W}{\langle \Phi_W \rangle} \right)^{1/3} \;\ge\; 3.4\times 10^{10}\,\text{GeV}\,.
\end{equation}
Assuming $\langle \Phi_W \rangle =f_W$, we find $\Lambda\gtrsim 3.4\times 10^{10}$ GeV.
Adopting a conservative, background-limited projection for Hyper-Kamiokande with
10~years of running assuming the sensitivity reached is $\tau_{p\to e^+ a_W} \simeq 2.1\times 10^{33}\ \text{yr}$ we find:
\begin{equation}
\Lambda \left( \frac{f_W}{\langle \Phi_W \rangle} \right)^{1/3} \;\ge\; 4.0\times 10^{10}\,\text{GeV}\,.
\end{equation}
We find that the constraint from $p\to e^{+}\pi^{0}$ is strictly stronger than that from $p\to e^{+}a_{W}$. Therefore, any parameter choice $(\Lambda,\,f_{W},\,\langle\Phi_{W}\rangle)$ that satisfies $p\to e^{+}\pi^{0}$ automatically satisfies $p\to e^{+}a_{W}$. The allowed parameter space is illustrated in Fig.~\ref{fig:aabound}.

The $U(1)_{B+L}$ charge of $\Phi_W$ also allows to add a lepton portal coupling
\begin{equation}
    \frac{\Phi_W}{\Lambda_l^2}\,(l H) (l H)\,,
\end{equation}
which contributes to Majorana neutrino masses (the scale $\Lambda_l$ may be different from the scale in Eq.~\eqref{eq:phiWop} as the lepton portal breaks $B\!-\!L$). As the absolute mass scale of neutrino masses is constrained by cosmological observations to be $m_\nu \lesssim 0.1$~eV \cite{DESI:2024mwx}, we find
\begin{equation}
  \Lambda_{\ell} \gtrsim \sqrt{\frac{\left\langle\Phi_W\right\rangle v^2}{m_\nu}}\approx 1.7 \times 10^{10}~\text{GeV}\,,
\end{equation}
where we assumed $\langle\Phi_W\rangle=10^6$ GeV. We note that this bound is still weaker than the one from proton decay.
Finally, a quartic portal with the Higgs, $\lambda_h\ \Phi_W^\dagger \Phi_W H^\dagger H$ would affect the Higgs coupling measurements. However, the lower limit on the $B\!+\!L$ breaking scale already constrain such effect to be $\lambda_h\ v_{\rm EW}/\langle \Phi_W \rangle \lesssim 10^{-4}$, hence below the experimental reach at current and future colliders \cite{deBlas:2019rxi}.

\section{Conclusions and outlook}
In this work we investigated a good quality weak axion, the $a_W$, realized as the pseudo--Nambu--Goldstone boson of a spontaneously broken $U(1)_{B+L}$ at the scale $f_W$. Through its $\mathrm{SU}(2)_L$ anomaly, $a_W$ dynamically relaxes the weak topological angle rendering the corresponding susceptibility unphysical. The anomaly coefficients, together with the tiny mass, make $a_W$ photophobic, while loops generate a log-enhanced coupling to fermions. 

The constraints on the $a_W$ parameter space are summarized in Fig.~\ref{fig:aabound}.
The loop-induced coupling to electrons allows stellar-cooling bounds from red giants to constrain $f_W \gtrsim 10^6\,\mathrm{GeV}$.
A distinctive feature of the $B{+}L$ construction is the portal $(\Phi_W/\Lambda^3)\,qqql$, which after symmetry breaking induces a dimension--six baryon--violating operator. Using Super--Kamiokande limits on $p\to e^+\pi^0$, we obtain
$\Lambda\gtrsim 8.3\times 10^{12}$~GeV for $f_W=10^6$~GeV. The alternative channel $p\to e^+ a_W$ yields a parametrically weaker constraint that is automatically satisfied whenever $p\to e^+\pi^0$ is satisfied. Looking ahead, Hyper--Kamiokande projections offer a mild improvement, pushing $\Lambda$ into the $10^{13}$~GeV range for $f_W\sim 10^6$~GeV. 
We also examined other portals. The lepton portal $(\Phi_W/\Lambda_\ell^2)(lH)(lH)$ contributes to Majorana neutrino masses and the upper bound on neutrino masses from cosmology requires $\Lambda_\ell \gtrsim  10^{10}~\text{GeV}$ for $f_W=10^6~\text{GeV}$. A quartic Higgs portal $\lambda_h\,\Phi_W^\dagger \Phi_W H^\dagger H$ is likewise innocuous in view of $v_{\rm EW}/\langle\Phi_W\rangle\lesssim 10^{-4}$. 

In summary, the $a_W$ is a \emph{good quality} solution and is essentially photophobic, but its baryon--violating portal makes proton decay the most incisive probe. Further improvements in stellar cooling systematics and proton decay searches offer the most promising avenues to test the $B{+}L$ weak axion paradigm.

\subsection*{Acknowledgments}
 The work of F.S. is partially supported by the Carlsberg Foundation, grant CF22-0922.

\bibliography{biblio}

\end{document}